\documentstyle [12pt,epsfig]{article}  
\catcode`@=11
\newcount\@tempcntc
\def\@citex[#1]#2{\if@filesw\immediate\write\@auxout{\string\citation{#2}}\fi
  \@tempcnta\z@\@tempcntb\m@ne\def\@citea{}\@cite{\@for\@citeb:=#2\do
    {\@ifundefined
       {b@\@citeb}{\@citeo\@tempcntb\m@ne\@citea\def\@citea{,}{\bf ?}\@warning
       {Citation `\@citeb' on page \thepage \space undefined}}%
    {\setbox\z@\hbox{\global\@tempcntc0\csname b@\@citeb\endcsname\relax}%
     \ifnum\@tempcntc=\z@ \@citeo\@tempcntb\m@ne
       \@citea\def\@citea{,}\hbox{\csname b@\@citeb\endcsname}%
     \else
      \advance\@tempcntb\@ne
      \ifnum\@tempcntb=\@tempcntc
      \else\advance\@tempcntb\m@ne\@citeo
      \@tempcnta\@tempcntc\@tempcntb\@tempcntc\fi\fi}}\@citeo}{#1}}
\def\@citeo{\ifnum\@tempcnta>\@tempcntb\else\@citea\def\@citea{,}%
  \ifnum\@tempcnta=\@tempcntb\the\@tempcnta\else
   {\advance\@tempcnta\@ne\ifnum\@tempcnta=\@tempcntb \else \def\@citea{--}\fi
    \advance\@tempcnta\m@ne\the\@tempcnta\@citea\the\@tempcntb}\fi\fi}
\catcode`@=12
\begin{document}

\newcommand{\gsim}{\;\raisebox{-0.9ex}{$\textstyle\stackrel{\textstyle>}{\sim}$}
 \;}
\newcommand{\lsim}{\;\raisebox{-0.9ex}{$\textstyle\stackrel{\textstyle<}{\sim}$}
 \;}

\begin{titlepage}

\pagenumbering{arabic}

%\begin{flushright}
%DELPHI 95-97 PHYS 532 \\
%revised version 5 Octobre, 1996
%\end{flushright}

\vspace*{1.0cm}
\begin{center}
\Large {\bf Multiplicity Fluctuations \\ in One- and
      Two-Dimensional Angular 
      Intervals \\ Compared with    
      Analytic QCD Calculations}\\

\vspace*{2.0cm}
\large {DELPHI Collaboration \\}
\vspace*{0.6cm}
\normalsize {B. Buschbeck and F. Mandl\\ Institute for High Energy Physics of the Austrian
Academy of Sciences, Vienna, Austria}
\vspace*{5.5cm}
\end{center}

\begin{abstract}
\noindent Multiplicity fluctuations in rings around the jet axis and in
off-axis cones have been measured by the DELPHI collaboration 
in $e^+e^-$ annihilations into hadrons 
at LEP energies. The measurements are compared with 
analytical perturbative QCD
calculations for the corresponding multiparton system, using the concept
of Local Parton Hadron Duality. Some qualitative features are confirmed 
by the data but substantial
quantitative deviations are observed. 
\end{abstract}
\clearpage

\end{titlepage}

\setcounter{page}{1}

\section{Introduction} 
To describe multiplicity fluctuations in angular regions by analytical
calculations using perturbative QCD is a challenge. It could help to improve
our understanding of the parton cascading mechanism and might lead
to a simple description of multiparticle correlations by QCD alone.
The idea that QCD jets might exhibit a self-similar (or fractal) structure 
was brought up already in 1979 by R.P.Feynman~\cite{feyn}, 
A.Giovannini~\cite{giov} and G.Veneziano~\cite{venez}.
In recent years this conception has been confirmed by various groups 
\cite{ochs,pesch,dremin}, giving detailed predictions on variables
and phase space regions where fractality is expected to show up. A simple
predicted dependence of the fractal dimensions on $\alpha_s$ 
stimulated further interest in measuring them experimentally. 

The analytical calculations 
are performed in the
Double Log Approximation (DLA)~\cite{parton,dla}, neglecting energy-momentum 
conservation, and
concern only idealized jets. 
They provide leading order predictions applicable quantitatively
at very high energies ( $\geq$ 1~TeV) \cite{ochs}. At LEP energies, 
non-perturbative effects may be important.
Also, they refer to multiparton states, whereas only 
multihadron states can be measured. 
It has been suggested that the parton evolution should be extended 
from the perturbative regime down to a lower mass scale (if possible to the
mass scale of light hadrons) to be able to compare the partonic states directly
with the hadronic states. 
This concept of Local Parton Hadron Duality
(LPHD) \cite{lhd1} is quite successful for single particle distributions 
and for global moments of multiplicity distributions. It remains 
questionable in the case of the more refined variables used here,
namely factorial moments and cumulants in phase space bins. 
First experimental
measurements~\cite{wien1,wien2,wien3,kittel,cone} revealed, indeed,
substantial deviations.

On the other hand it can be expected that these calculations will
improve in the future. This would provide us with a better understanding of the
internal structure of jets in terms of analytical expressions than 
can be obtained by Monte Carlo calculations with many
parameters. In fact, the analytical predictions considered in this paper
involve only one adjustable parameter, namely the QCD scale $\Lambda$.

 The aim of this study is to use DELPHI data to measure 
multiplicity fluctuations in one- and
two-dimensional angular intervals and compare them with the available 
theoretical predictions. It is hoped that such a study may 
show how to approach nearer to a satisfying theory based on QCD and LPHD 
which describes high energy multiparticle phenomena.

In section 2 the theoretical framework is sketched,
section 3 contains information about the experimental data and the
Monte Carlo comparisons and in section 4 the comparison with the
analytical calculations is presented. Section 5 contains the final 
discussion and the
summary.    

\section{Theoretical framework} 
The theoretical calculations treat correlations between partons 
emitted within an angular window 
%of width $\vartheta$ positioned at a polar angle $\Theta$ 
defined by two angles $\vartheta$ and $\Theta$. 
%(well described in~\cite{dremin}).  
The parton and particle density  
correlations (fluctuations) in this window are described by 
normalized factorial moments of order $n$:
% and and by their dependence on the 
%window size $\vartheta$:
\begin{equation}
F^{(n)}(\Theta ,\vartheta )  =  \frac{\int 
\rho^{(n)} (\Omega_1 , \ldots , 
\Omega_n)  d\Omega_1 \ldots d\Omega_n }
{\int
\rho^{(1)}(\Omega_1) \ldots \rho^{(1)}(\Omega_n) d\Omega_1 \ldots d\Omega_n} 
\end{equation}
where $\rho^{(n)} (\Omega_1 , \ldots , \Omega_n)$ are the $n$-parton/particle 
density 
correlation functions which depend on the spherical angles 
$\Omega_k$. The integrals extend over the window chosen. 

The angular windows considered here are either rings around the 
jet axis with mean 
opening angle $\Theta = 25^\circ$ and half width $\vartheta$ in the 
case of 1 dimension ($D=1$), or cones with half opening angle $\vartheta$ around
a direction ($\Theta,\Phi$) with respect to the jet axis in the case of
2 dimensions ($D=2$). At sufficiently large jet energies, the parton flow in 
these angular windows is dominated by parton avalanches caused
by gluon bremsstrahlung off the initial quark.

The cumulants $C^{(n)}$ are obtained from the moments $F^{(n)}$
by simple algebraic equations~\cite{cumul11}, e.g. $C^{(2)} = F^{(2)} - 1$,
$C^{(3)} = F^{(3)} - 3(F^{(2)} - 1) -1$ .

The theoretical scheme for deriving the moments described above 
is based on the generating functional techniques \cite{dla,gen10} 
in the DLA of perturbative QCD. The
probability of radiating a gluon with momentum $k$ at an emission angle
$\Theta_g$ and azimuthal angle $\Phi_g$ from an initial parton $a$ has 
been approximated by
\begin{equation}
  M(k)d^3k = c_a \gamma_0^2 \frac{dk}{k}\frac{d\Theta_g}{\Theta_g}\frac{d\Phi_g}
{2\pi}
\end{equation}
\begin{equation}
   \gamma_0^2 = 6\alpha_S/\pi 
\end{equation}
with $c_a$ = 1 if $a$ is a gluon and $c_a$ = 4/9 if $a$ is a quark.

Ref.~\cite{ochs} derived their predictions explicitly for cumulant moments
$C^{(n)}$,
whereas~\cite{pesch} and~\cite{dremin} obtained similar expressions for the
factorial moments $F^{(n)}$. It has been shown~\cite{ochs} by Monte Carlo 
calculations that,
at very high energy ($\sqrt{s} \geq 1800$~GeV), the values of 
$F^{(n)}$ and $C^{(n)}$ converge to each other. 
At LEP energies, however, the cumulants are still
far away from the asymptotic predictions (see section 4). 

For the normalized cumulant moments
$C^{(n)}$~\cite{ochs} and the factorial moments
$F^{(n)}$~\cite{pesch,dremin},
the following prediction  has been made:
\begin{equation} C^{(n)}(\Theta,\vartheta )~\mbox{or}~F^{(n)} (\Theta ,\vartheta ) \sim 
\left(\frac{\Theta}{\vartheta}\right)^{\phi_n}
\end{equation}
All 3 references~\cite{ochs,pesch,dremin} give in the high energy limit 
and for large values of $\vartheta \leq
 \Theta$ the same linear
approximation for the exponents $\phi_n$:
\begin{equation}
\phi_n \approx (n-1)D - \left(n - \frac{1}{n}\right) \gamma_0 
\end{equation}
\noindent where D is a dimensional factor, 1 for ring regions and 2 for cones.
For fixed $\alpha_s$ (along the parton shower)  eq.~5 is 
asymptotically
valid for \underline{all} angles. In this case
the fractal (Renyi-) dimension $D_n$~\cite{ren}
can be obtained~\cite{lip} from $\phi_n$ (eq.~5) via:

\begin{eqnarray}
D_n = D - \frac{\phi_n}{n-1} \\
D_n = \frac{n+1}{n}\gamma_0
\end{eqnarray}

When the running of $\alpha_S$ with $\vartheta$ in the parton cascade is taken 
into
account, in \cite{ochs} the following was obtained

\begin{equation}
\phi_n \approx (n-1)D - 2\gamma_0 (n-\omega (\epsilon , n))/\epsilon
\end{equation}

\begin{equation} 
\omega(\epsilon ,n) = n\sqrt{1-\epsilon}(1-\frac{1}{2n^2} \ln(1-\epsilon))
\end{equation}     
and
\begin{equation}
\epsilon = \frac{\ln(\Theta/\vartheta)}{\ln(P\Theta/\Lambda)}
\end{equation}
where $P \approx 
\sqrt{s}/2$ is the momentum of the initial parton.
The dependence on the QCD parameters $\alpha_s$ or $\Lambda$ enters in the
above equations via $\gamma_0$ and $\epsilon$ that are determined by the scale
$Q \approx P\Theta$.
In the present study it is about 20 GeV for $\sqrt{s}$=91.1~GeV.

The corresponding predictions of refs.~\cite{pesch} (eq.~11) and~\cite{dremin}
(eq.~12) are analytically different, but numerically similar:

\begin{equation}
\phi_n = (n-1)D - \frac{2\gamma_0}{\epsilon} \cdot \frac{n^2 -1}{n} \left(1 -
\sqrt{1-\epsilon}\right)
\end{equation}

\begin{equation}
\phi_n = (n-1)D - \frac{n^2 - 1}{n} \gamma_0 \left( 1 + \frac{n^2 +1} {4 n^2}
\epsilon\right)
\end{equation}

It should be noted that all three theoretical papers cited above use the 
lowest order QCD relation (13) between the coupling $\alpha_s$ and the
QCD scale $\Lambda$, which is also used in the present analysis:
\begin{equation}
\alpha_s = \frac{\pi \beta^2}{6} \frac{1}{\ln(Q/\Lambda)}
\end{equation}
\begin{equation}
\beta^2 = 12\left(\frac{11}{3}n_c - \frac{2}{3}n_f\right)^{-1}
\end{equation}
where $n_c$=3 (number of colours).
These relations depend also on the number of flavours ($n_f$). Since
eq.~13 emerges only from ``one loop'' calculations, the parameter $\Lambda$
is not the universal $\Lambda_{\overline{MS}}$, but only
an effective parameter $\Lambda_{eff}$.
But also in this approximation $\alpha_s$ runs, having a
scale dependence $1/\ln(Q^2/\Lambda^2)$.

The running of $\alpha_s$ during the process of jet cascading is 
implicitly taken into account in (8), (11) and (12) by the dependence of
$\phi_n$ on $\epsilon$ (or $\vartheta$). In theory this causes a 
deviation from a potential behaviour (eqs.~4 and 5) 
of $F^{(n)}$ 
when approaching smaller values of $\vartheta$ (larger $\epsilon$).

All theoretical predictions concern the partonic states. The
corresponding experimental measurements, however, are of
hadronic states. When comparing them, the hypothesis of LPHD has to be
used. 

It may be noted 
that the factorial moments $F^{(n)}$ measured in the present study (see 
also eq.~15 below) are very similar to the
well known and previously measured moments in rapidity space. Here  
the angle $\vartheta$ is used (translated by constant factors into
$\epsilon$), because this is the natural variable in the QCD calculations.

\section{Experimental data and comparison with the Monte Carlo calculations}

The normalized factorial moments (1) are determined experimentally 
by counting $n_m$, the
number of charged particles in the respective windows of phase 
space, for each event:
\begin{equation}
F^{(n)}(\Theta ,\vartheta )  =  \frac{\left< n_m (n_m -1) 
\ldots (n_m - n+1)\right>} 
{\left< n_m\right>^n}
\end{equation}
where the brackets $< >$ denote averages over the whole event sample.

The data sample used contains about 600000 $e^+e^-$ interactions (after cuts) 
collected by DELPHI  at $\sqrt{s} = 91.1$~GeV in 1994. 
A sample of about 1200 high energy events at $\sqrt{s}$=183~GeV incident
energy collected in
1997 is used to investigate the energy dependence. The calculated hadron
energy was required to be greater than 162~GeV (corresponding to
a mean energy of 175~GeV).
The standard cuts as in \cite{delphi} for hadronic events and track quality 
were applied by demanding a minimum charged multiplicity, enough visible
charged energy and events well contained within the detector volume. In the
present study all charged particles (except identified electrons and
muons) with momentum larger than 0.1~GeV have
been considered.
The special procedures for selecting high energy events are described in
\cite{high}.
WW-events have been 
excluded.
Detailed Monte Carlo studies were done 
using the JETSET 7.4 PS model~\cite{lund}.
 The corrections were determined using events 
from a JETSET Monte Carlo simulation which had
been tuned ($\Lambda$=0.346~GeV and $Q_0$=2.25~GeV)
to reproduce general event characteristics~\cite{ham}, which 
included variables different from those referred to in section 2. 
These
events were examined at
\begin{itemize}
\item[*] {\bf Generator level}\\ where all charged final--state particles 
(except electrons and muons) with a mean lifetime
longer than $10^{-9}$ seconds have been considered;

\item[*] {\bf Detector level}\\ which includes distortions due to particle
decays and interactions with the detector material, other imperfections such as 
limited resolution, multi--track separation and
detector acceptance, and the event selection procedures.
\end{itemize}

Using these events, the factorial moments and cumulants introduced in
section 2 of order $n$, 
$A_n$, studied below were corrected
(for each $\epsilon$ interval considered) by

\begin{equation} A^{\rm{cor}}_n \, = \, g_n A^{\rm{raw}}_n \, , \quad g_n \, = \,
\frac{A^{\rm{gen}}_n}{A^{\rm{det}}_n}
\end{equation}

\noindent where the superscript ``raw'' indicates the quantities 
calculated directly
from the data, and ``gen'' and ``det'' denote those obtained from the 
Monte Carlo events at
generator and detector level respectively. 
The simulated data at detector level were found to agree satisfactorily 
with the
experimental data. The
measurement error on the relative angle $\vartheta_{12}$ between two 
outgoing particles
was determined to be of order $0.5^\circ$ 
(if both tracks had good Vertex Detector hits, even as small as $0.1^\circ$).
The jet axis is chosen to be the sphericity axis.
To increase statistics in the case of the high energy sample the moments
(15) have been calculated in both sphericity hemispheres and averaged.

In addition, all phenomena which were not included in the analytical 
calculations had to
be corrected for, namely (i) initial state photon radiation,
(ii) Dalitz decays of the $\pi^0$, 
(iii) residual $K^0_s$ and $\Lambda^0$ decays near the vertex, 
and (iv) the effect of 
Bose--Einstein correlations.
The corrections were estimated, for each $\epsilon$ interval, 
like $g_n$ in eq.~16, 
by switching the effects on and off. Each of these correction factors 
were found to be 
below 10\% in the case of factorial moments. 
The largest corrections have been found 
in the case of cumulants of higher orders and amounted to 16--25\%,
depending on the analysis angle. 

The total correction factor including all
effects is denoted by $g^{\rm{tot}}_n$ and is the product of the 
individual factors. Systematic errors have 
been calculated from
$g^{\rm{tot}}_n$ according to $\Delta A^{\rm{corr}}_n = \pm | A^{\rm{raw}}_n
(g^{\rm{tot}}_n - 1)/2|$. Due to uncertainties in measuring multiple 
tracks at very
small separation angles, an additional systematic error was added for small
$\vartheta$ values for $F^{(4)}$ and $F^{(5)}$.

\begin{figure}
\vspace{-2cm}
\mbox{\epsfig{file=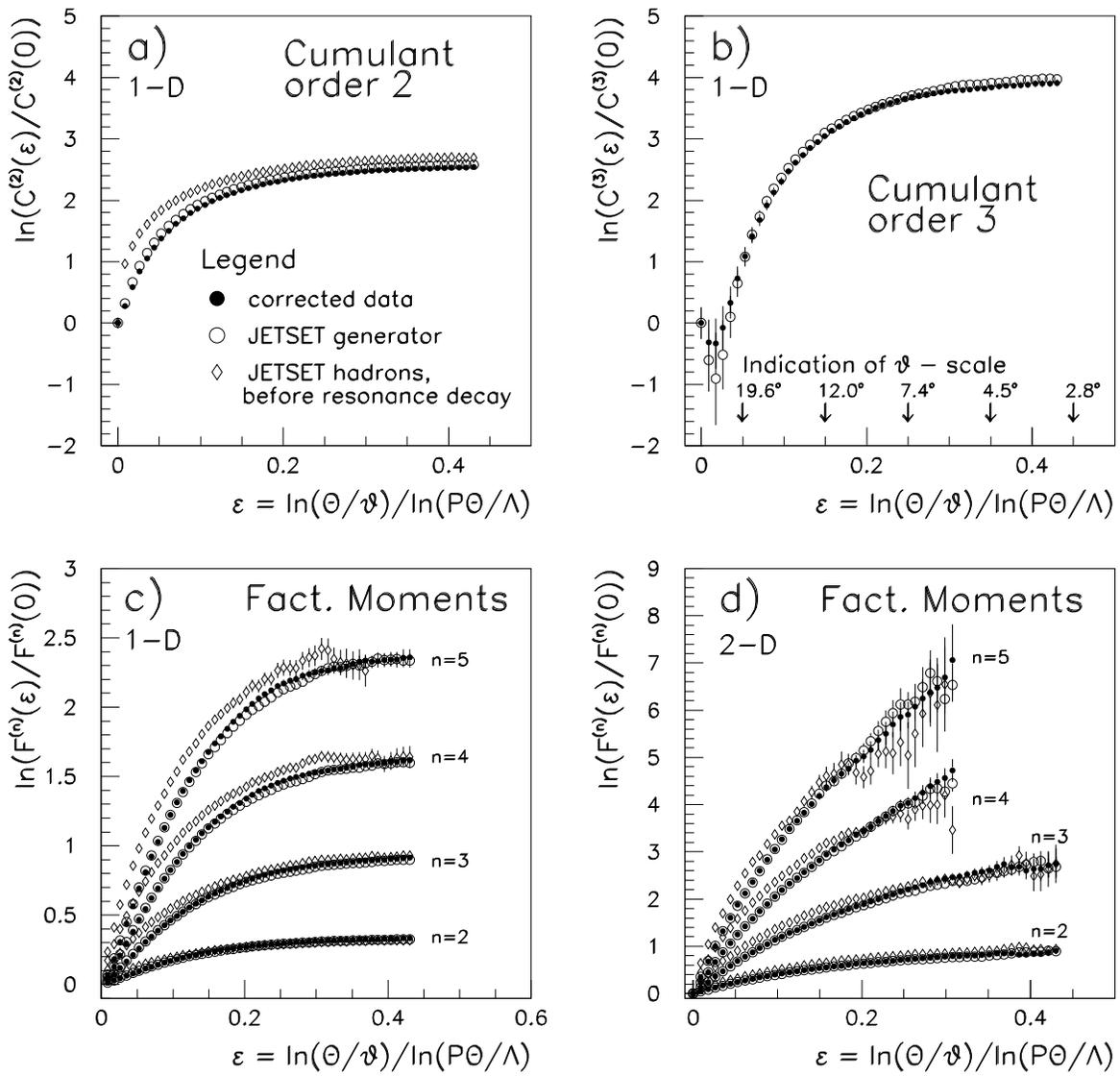,width=17cm}}
\vspace{-1.9cm}
\caption[]{ 
\footnotesize 
{\bf a)} The measured ($\bullet$) cumulants of order $n=2$
in 1-D rings, {\bf b)}
those of order $n=3$, {\bf c)} the factorial moments of orders $n=2$ to 5 in
1-D rings, and {\bf d)} those in 2-D off-axis cones are 
compared with 
JETSET 7.4 before ($\diamond$) and after ($\circ$) resonance decay. The polar
angle $\Theta = 25^\circ$. Only charged hadrons have been considered. Because of
a negative $C^3(0)$ value in JETSET before resonance decay, no normalization 
was possible in this case and the corresponding points have therefore been 
omitted in {\bf b)}. Values of $\vartheta$ which 
correspond to the respective values of $\epsilon$, with $\Lambda = 0.15$~GeV, 
are also indicated in {\bf b)}. 
}
\end{figure}

Fig.~1 shows a comparison 
at $\sqrt{s}$=91.1~GeV 
of the measured 1-dimensionel cumulants 
and 1- and
2-dimensional factorial moments with JETSET 7.4 tuned as described above.
The cumulants and factorial moments are 
normalized by $C^{(n)}(0)$ and $F^{(n)}(0)$ for easy comparison of the
measured shapes with the analytical predictions. 
There is generally good agreement between the
Monte Carlo simulation (open circles) and the corrected data (full circles).
The study of the influence of the resonance decay shown in Fig.~1
reveals significant effects.
Numerical values of the measured and corrected 1- and 2-dimensional 
factorial moments
are given in Tables 1 and 2, respectively, for convenience as function of
$\vartheta$/$\Theta$ (the $\epsilon$ dependence follows from eq.~10).

\begin{center}
{{\bf Table 1a : 1-dimensional factorial moments for orders 2 and 3}\\
{\small together with their statistical and systematic errors
as function of $\vartheta$/$\Theta$ ($\Theta = 25^\circ$)}}\\
\vspace{3mm}
%\begin{scriptsize}
\begin{tabular}{|c|c|c|c|c|c|c|}
\hline
{\bf $\vartheta/\Theta$} & {\bf $F_2$} & {\bf $\pm$ stat}
& {\bf $\pm$ syst} & {\bf $F_3$} & {\bf $\pm$ stat} &
{\bf $\pm$ syst}  \\
\hline & & & & & & \\
1.0000 & 1.035 & 0.002 & 0.009 & 1.114 & 0.003 & 0.030 \\
0.9180 & 1.063 & 0.002 & 0.009 & 1.196 & 0.003 & 0.031 \\
0.8426 & 1.101 & 0.002 & 0.010 & 1.315 & 0.004 & 0.034 \\
0.7735 & 1.139 & 0.002 & 0.010 & 1.446 & 0.005 & 0.037 \\
0.7100 & 1.176 & 0.003 & 0.010 & 1.577 & 0.006 & 0.040 \\
0.6518 & 1.210 & 0.003 & 0.011 & 1.707 & 0.007 & 0.043 \\
0.5983 & 1.241 & 0.003 & 0.011 & 1.831 & 0.008 & 0.045 \\
0.5492 & 1.269 & 0.003 & 0.012 & 1.949 & 0.009 & 0.047 \\
0.5041 & 1.293 & 0.004 & 0.012 & 2.055 & 0.010 & 0.048 \\
0.4628 & 1.316 & 0.004 & 0.013 & 2.156 & 0.012 & 0.050 \\
0.4248 & 1.335 & 0.004 & 0.013 & 2.246 & 0.013 & 0.051 \\
0.3899 & 1.353 & 0.004 & 0.014 & 2.330 & 0.014 & 0.054 \\
0.3579 & 1.368 & 0.005 & 0.015 & 2.406 & 0.015 & 0.057 \\
0.3286 & 1.382 & 0.005 & 0.015 & 2.474 & 0.017 & 0.060 \\
0.3016 & 1.393 & 0.005 & 0.016 & 2.530 & 0.018 & 0.061 \\
0.2769 & 1.402 & 0.005 & 0.016 & 2.576 & 0.019 & 0.064 \\
0.2541 & 1.409 & 0.006 & 0.017 & 2.612 & 0.021 & 0.067 \\
0.2333 & 1.416 & 0.006 & 0.018 & 2.646 & 0.022 & 0.071 \\
0.2141 & 1.422 & 0.006 & 0.018 & 2.672 & 0.024 & 0.073 \\
0.1966 & 1.428 & 0.006 & 0.019 & 2.699 & 0.025 & 0.079 \\
0.1804 & 1.432 & 0.007 & 0.020 & 2.720 & 0.027 & 0.086 \\
0.1656 & 1.435 & 0.007 & 0.020 & 2.740 & 0.029 & 0.092 \\
0.1520 & 1.437 & 0.007 & 0.020 & 2.751 & 0.031 & 0.094 \\
0.1396 & 1.441 & 0.008 & 0.021 & 2.771 & 0.033 & 0.097 \\
0.1281 & 1.444 & 0.008 & 0.022 & 2.785 & 0.035 & 0.104 \\
0.1176 & 1.448 & 0.008 & 0.022 & 2.797 & 0.037 & 0.109 \\
0.1080 & 1.451 & 0.009 & 0.022 & 2.808 & 0.040 & 0.108 \\
0.0991 & 1.454 & 0.009 & 0.022 & 2.812 & 0.043 & 0.106 \\
 & & & & & &\\
\hline
\end{tabular}
%\end{scriptsize}
\end{center}
\vfill
\begin{center}
{\bf Table 1b : 1-dimensional factorial moments for orders 4 and 5 )}\\
{\small together with their statistical and systematic errors
as function of $\vartheta/\Theta$} ($\Theta=25^{\circ}$)\\
\vspace{3mm}
%\begin{scriptsize}
\begin{tabular}{|c|c|c|c|c|c|c|}
\hline
{\bf $\vartheta/\Theta$} & {\bf $F_4$} & {\bf $\pm$ stat}
& {\bf $\pm$ syst} & {\bf $F_5$} & {\bf $\pm$ stat} &
{\bf $\pm$ syst}  \\
\hline & & & & & & \\
1.0000 & 1.251 & 0.005 & 0.068 &  1.465 & 0.010 & 0.135 \\
0.9180 & 1.417 & 0.006 & 0.076 &  1.757 & 0.013 & 0.160 \\
0.8426 & 1.681 & 0.008 & 0.089 &  2.270 & 0.019 & 0.203 \\
0.7735 & 1.994 & 0.011 & 0.103 &  2.931 & 0.028 & 0.253 \\
0.7100 & 2.332 & 0.015 & 0.116 &  3.697 & 0.039 & 0.307 \\
0.6518 & 2.689 & 0.018 & 0.130 &  4.568 & 0.052 & 0.362 \\
0.5983 & 3.047 & 0.022 & 0.141 &  5.489 & 0.067 & 0.412 \\
0.5492 & 3.406 & 0.027 & 0.152 &  6.467 & 0.086 & 0.461 \\
0.5041 & 3.745 & 0.032 & 0.157 &  7.440 & 0.106 & 0.491 \\
0.4628 & 4.085 & 0.038 & 0.165 &  8.467 & 0.130 & 0.528 \\
0.4248 & 4.395 & 0.043 & 0.173 &  9.436 & 0.158 & 0.573 \\
0.3899 & 4.694 & 0.050 & 0.186 & 10.403 & 0.190 & 0.634 \\
0.3579 & 4.974 & 0.056 & 0.200 & 11.337 & 0.220 & 0.837 \\
0.3286 & 5.227 & 0.063 & 0.211 & 12.196 & 0.257 & 1.200 \\
0.3016 & 5.438 & 0.071 & 0.262 & 12.931 & 0.304 & 1.501 \\
0.2769 & 5.605 & 0.079 & 0.260 & 13.464 & 0.347 & 1.320 \\
0.2541 & 5.739 & 0.087 & 0.261 & 13.924 & 0.398 & 1.508 \\
0.2333 & 5.855 & 0.096 & 0.261 & 14.286 & 0.456 & 1.508 \\
0.2141 & 5.939 & 0.105 & 0.263 & 14.520 & 0.508 & 1.503 \\
0.1966 & 6.021 & 0.112 & 0.299 & 14.714 & 0.536 & 1.521 \\
0.1804 & 6.113 & 0.123 & 0.347 & 15.125 & 0.611 & 1.543 \\
0.1656 & 6.194 & 0.134 & 0.395 & 15.447 & 0.679 & 1.671 \\
0.1520 & 6.228 & 0.147 & 0.402 & 15.381 & 0.748 & 1.633 \\
0.1396 & 6.292 & 0.160 & 0.424 & 15.493 & 0.819 & 1.742 \\
0.1281 & 6.335 & 0.176 & 0.467 & 15.610 & 0.942 & 1.959 \\
0.1176 & 6.351 & 0.189 & 0.510 & 15.544 & 1.005 & 2.333 \\
0.1080 & 6.361 & 0.207 & 0.512 & 15.485 & 1.091 & 2.438 \\
0.0991 & 6.289 & 0.219 & 0.502 & 14.637 & 1.122 & 2.279 \\
 & & & & & &\\
\hline
\end{tabular}
%\end{scriptsize}
\end{center}
\newpage

\begin{center}
{\bf Table 2a : 2-dimensional factorial moments for orders 2 and 3}
\\{\small together with their statistical and systematic errors
as function of $\vartheta/\Theta$ ($\Theta=25^{\circ}$)}\\
\vspace{3mm}
%\begin{scriptsize}
\begin{tabular}{|c|c|c|c|c|c|c|}
\hline
{\bf $\vartheta/\Theta$} & {\bf $F_2$} & {\bf $\pm$ stat}
& {\bf $\pm$ syst} & {\bf $F_3$} & {\bf $\pm$ stat} &
{\bf $\pm$ syst}  \\
\hline & & & & & & \\
  1.000 &   1.046 &   0.002 &   0.036 &   1.155 &   0.004 &   0.111 \\
  0.918 &   1.143 &   0.002 &   0.041 &   1.476 &   0.006 &   0.145 \\
  0.843 &   1.240 &   0.003 &   0.047 &   1.858 &   0.010 &   0.193 \\
  0.774 &   1.337 &   0.004 &   0.056 &   2.296 &   0.015 &   0.258 \\
  0.710 &   1.428 &   0.005 &   0.067 &   2.769 &   0.022 &   0.338 \\
  0.652 &   1.518 &   0.006 &   0.080 &   3.298 &   0.031 &   0.435 \\
  0.598 &   1.602 &   0.007 &   0.095 &   3.863 &   0.043 &   0.544 \\
  0.549 &   1.678 &   0.009 &   0.112 &   4.440 &   0.058 &   0.660 \\
  0.504 &   1.749 &   0.010 &   0.129 &   5.085 &   0.077 &   0.784 \\
  0.463 &   1.822 &   0.012 &   0.148 &   5.821 &   0.103 &   0.914 \\
  0.425 &   1.888 &   0.014 &   0.167 &   6.506 &   0.132 &   1.021 \\
  0.390 &   1.956 &   0.017 &   0.186 &   7.312 &   0.171 &   1.123 \\
  0.358 &   2.011 &   0.019 &   0.204 &   8.067 &   0.222 &   1.186 \\
  0.329 &   2.069 &   0.022 &   0.221 &   9.007 &   0.298 &   1.240 \\
  0.302 &   2.121 &   0.026 &   0.236 &   9.994 &   0.398 &   1.258 \\
  0.277 &   2.188 &   0.030 &   0.251 &  10.985 &   0.503 &   1.233 \\
 & & & & & &\\
\hline
\end{tabular}
%\end{scriptsize}
\end{center}
\begin{center}
{\bf Table 2b : 2-dimensional factorial moments for orders 4 and 5}
\\{\small together with their statistical and systematic errors
as function of $\vartheta/\Theta$ ($\Theta = 25^{\circ}$)}\\
\vspace{3mm}
%\begin{scriptsize}
\begin{tabular}{|c|c|c|c|c|c|c|}
\hline
{\bf $\vartheta/\Theta$} & {\bf $F_4$} & {\bf $\pm$ stat}
& {\bf $\pm$ syst} & {\bf $F_5$} & {\bf $\pm$ stat} &
{\bf $\pm$ syst}  \\
\hline & & & & & & \\
  1.000 &   1.356 &   0.008 &   0.250 &   1.703 &   0.021 &   0.508 \\
  0.918 &   2.126 &   0.018 &   0.397 &   3.369 &   0.056 &   1.009 \\
  0.843 &   3.234 &   0.035 &   0.628 &   6.343 &   0.138 &   1.913 \\
  0.774 &   4.738 &   0.063 &   0.963 &  11.133 &   0.293 &   3.368 \\
  0.710 &   6.614 &   0.107 &   1.405 &  18.072 &   0.575 &   5.428 \\
  0.652 &   9.046 &   0.182 &   1.988 &  28.813 &   1.187 &   8.458 \\
  0.598 &  12.110 &   0.297 &   2.706 &  44.913 &   2.309 &  12.622 \\
  0.549 &  15.604 &   0.467 &   3.469 &  65.771 &   4.404 &  17.235 \\
  0.504 &  20.094 &   0.693 &   4.326 &  94.370 &   6.777 &  22.310 \\
  0.463 &  25.785 &   1.050 &   5.206 & 134.110 &  11.377 &  27.410 \\
  0.425 &  31.593 &   1.505 &   5.758 & 179.450 &  17.935 &  29.940 \\
  0.390 &  38.817 &   2.178 &   6.087 & 234.930 &  27.941 &  29.300 \\
  0.358 &  46.590 &   3.379 &   5.895 & 297.580 &  53.980 &  23.620 \\
  0.329 &  58.358 &   5.432 &   5.405 & 419.930 & 104.740 &  13.500 \\
  0.302 &  72.636 &   8.664 &   4.098 & 599.320 & 181.600 &  16.410 \\
  0.277 &  85.249 &  11.582 &   1.647 & 743.900 & 235.130 &  46.850 \\
 & & & & & &\\
\hline

\end{tabular}
%\end{scriptsize}
\end{center}

\begin{figure}
\mbox{\epsfig{file=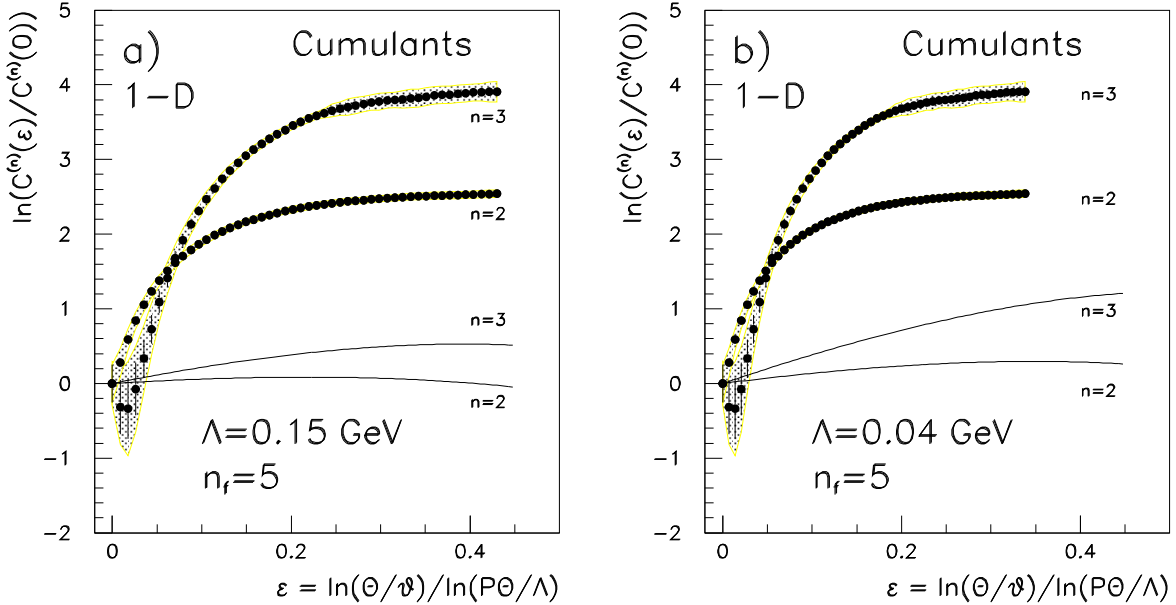,width=17cm}}
\vspace{-1.9cm}
\caption[]{ 
\footnotesize 
The second and third order cumulants (full circles) are compared 
with the predictions~\cite{ochs}, eq.~8 (solid lines)
with $n_f = 5$ for {\em a)} $\Lambda = 0.15$~GeV, {\em b)}
$\Lambda = 0.04$~GeV. The statistical errors are shown by the error bars,
the systematic errors by the shaded regions.}
\end{figure}

\begin{figure}
\mbox{\epsfig{file=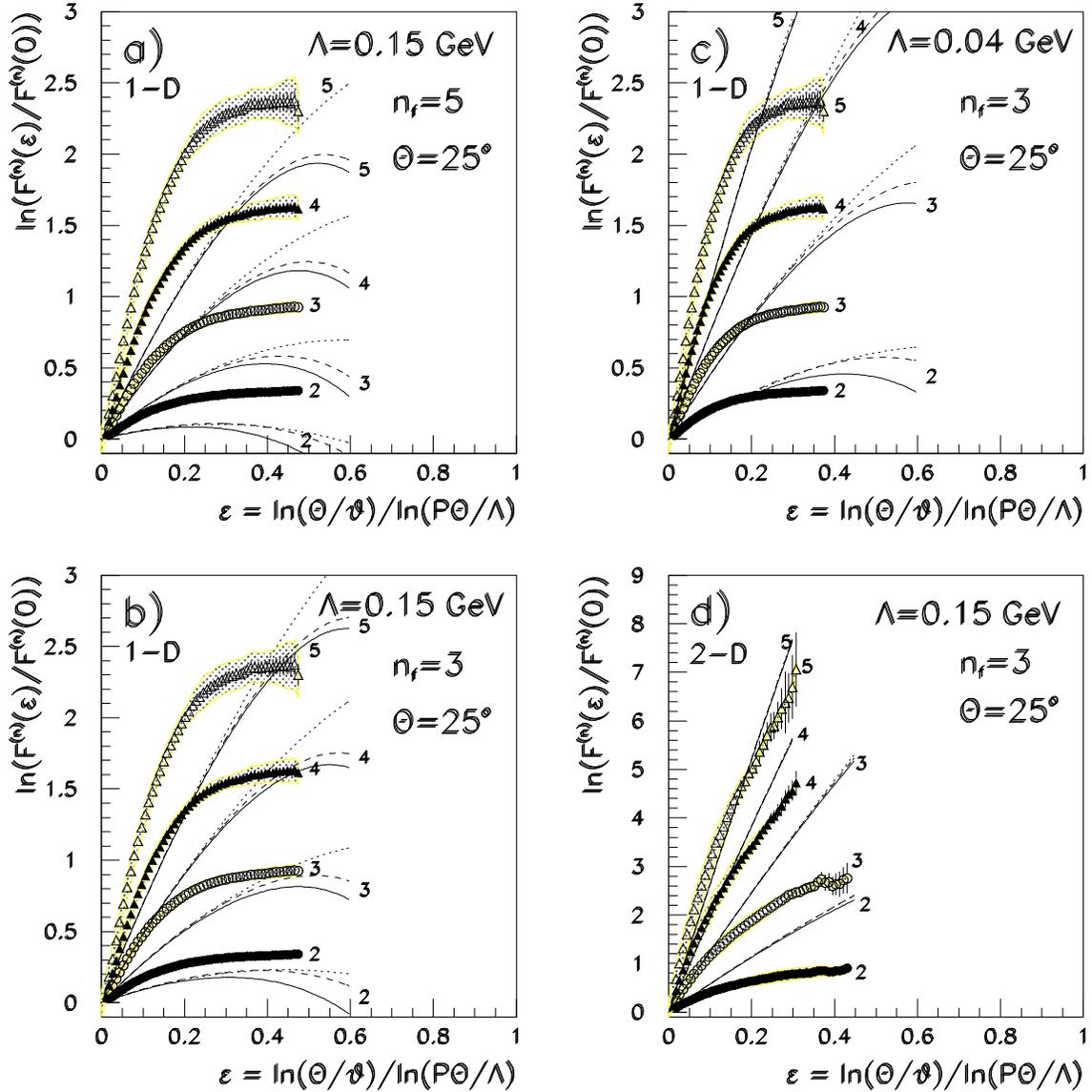,width=17cm}}
\vspace{-1.5cm}
\caption[]{
\footnotesize 
Factorial moments in 1-dimensional rings are compared with the 
analytical calculations
of refs~\cite{ochs,pesch,dremin}, eqs.~8 (solid lines), 11 (dashed lines),
and 12 (dotted lines). The dependences on $n_f$ and $\Lambda$ are shown in
a), b) and c). As a consistency test, 1- and 2-dimensional factorial moments are
compared in b) and d) with same QCD parameters: note the different vertical 
scales. The  
orders 2 to 5 are indicated in all figures, the data are also distinguished
by different symbols.
The statistical errors are shown by the error bars,
the shaded regions indicate the systematic errors.
The 1-dimensional factorial moments agree very well with those measured
by L3 \cite{kittel}.
}
\end{figure}

\section{ Comparison with the analytical calculations }

\subsection{Quantitative comparison at $\sqrt{s}$ = 91.1~GeV}
Fig.~2 shows the cumulants of orders $n$ = 2 and $n$ = 3 
in one-dimensional rings around jet cones
normalized by $C^{(n)}(0)$ 
and compared with the predictions of ref.~\cite{ochs}.
\begin{itemize}
\item[*]
The agreement with the data is very
bad: The predictions lie well below the data and differ in shape (Fig.~2a). 
Using a lower value of $\Lambda$ 
(i.e. $\Lambda = 0.04$~GeV instead of 0.15~GeV) does not help, 
as can be seen in Fig.~2b (neither does a smaller value of $n_f$, not shown 
here). 
\end{itemize} 

Fig.~3 shows the factorial moments of orders 2, 3, 4 and 5 normalized by 
$F^{(n)}(0)$, together with the predictions of refs.~\cite{ochs,pesch,dremin}, 
in one- and two- dimensional angular intervals (i.e. rings and side cones) for
various numerical values of $\Lambda$ and $n_f$.

\begin{itemize}
\item[*] The correlations in one-dimensional rings around jets, 
expressed by factorial moments, 
are not described well by the theoretical predictions~\cite{ochs,pesch,dremin}
using the QCD parameters $\Lambda = 0.15$~GeV and $n_f = 5$ (Fig.~3a). 
The predictions lie below the data for not too large $\epsilon$, 
differing also in shape. 
%Especially the higher moments $n \geq 3$ show large discrepancies.
\item[*]Choosing $n_f = 3$ (Fig.~3b) instead of $n_f = 5$ 
as in Fig.~3a reduces the discrepancies. 
\item[*] Choosing in addition the smaller value of $\Lambda = 0.04$~GeV 
(Fig.~3c), $F^{(2)}$ is well predicted for smaller
values of $\epsilon$,  
the higher orders ($n > 2$) still deviate considerably. 
%and there remain strong discrepancies for $n > 2$.   
\item[*]The factorial moments in 1 and 2 dimensions show different
behaviour for the lower order moments $n < 4$ : choosing the same set of
parameters ($\Lambda = 0.15$~GeV, $n_f = 3$),  
$F^{(2)}$ and $F^{(3)}$ lie \underline{above} the predictions
in the 1-dimensional case (Fig.~3b), 
but \underline{below} them in the 2-dimensional case (Fig.~3d).
\item[*]The higher moments $F^{(4)}$ and $F^{(5)}$ 
have similar features in the 1- and 2-dimensional case
(Figs.~3b,3d).
\item[*]In Fig.~3  
the slopes at small $\epsilon$ are generally steeper than
predicted (with the exception of $F^{(2)}$ and $F^{(3)}$
in Fig.~3d)
and the ``bending'' begins at smaller values of $\epsilon$.
\item[*]It is not possible to find \underline{one} set of QCD parameters 
$\Lambda$ and $n_f$
which simultaneously minimize the discrepancies between data and predictions 
for moments
of all orders 2,3,4 and 5 in both the 1- and 2-dimensional cases.  
\end{itemize}
\subsection{Energy dependence}
Fig.4 shows a comparison with high energy data at $\sqrt{s}$=183~GeV 
(with a mean energy of $\sqrt{s}$=175~GeV) and the
corresponding predictions according to eq.~8, where the energy dependence
enters via the parameter $\gamma_0$. 
It can be seen that for small values of $\epsilon$ there is no improvement 
of agreement at high energy. For larger values of $\epsilon$ the statistical
errors of the high energy data are substantial. The relative increase of the
predicted moments agrees qualitatively with that of the JETSET model that,
as shown in Fig.~1, agrees very well with the measurement at 
$\sqrt{s}$=91.1~GeV.
Similar conclusions
can be found from the predictions based on eqs.~11 and 12. 
\begin{figure}
\vspace{-2cm}
\mbox{\epsfig{file=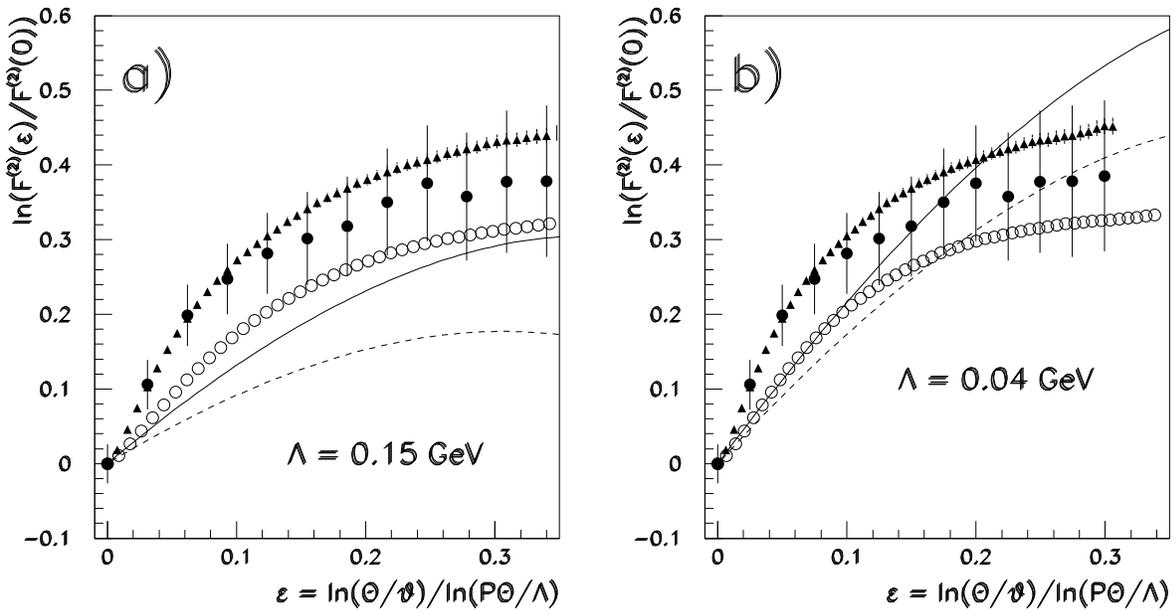,width=17cm}}
\vspace{-1.9cm}
\caption[]{ 
\footnotesize 
The energy dependence of the normalized factorial moment of order 2. 
{\bf $\sqrt{s}$=91.1~GeV}: data (open circles) and prediction ref.[4] 
dashed lines and
{\bf $\sqrt{s}$=175~GeV}: data (full circles) and prediction ref.[4] 
solid lines,  
for the QCD-parameter combinations {\bf a)} ($n_f=3$, $\Lambda$=0.15) and 
{\bf b)} ($n_f$=3, $\Lambda$=0.04).
The full triangles denote the high energy JETSET simulation. 
}
\end{figure}

\subsection{Qualitative features}
In the introduction, arguments have been given that the DLA might not be 
accurate enough for a quantitative description of experiments. 
Some disagreement with the measurement could be expected 
considering the asymptotic nature of the calculations, 
but nevertheless
an overall qualitative description of the data should be provided. 
Indeed 
the data (see Figs.~3,4) show some general qualitative features that are
predicted well by the analytical calculations:
\begin{itemize}
\item[*]The factorial moments rise linearly at small
$\epsilon$ exhibiting a fractal structure as predicted in eqs.~4,5 
for the parton cascade 
and saturate at higher values.
\item[*]The factorial moments increase from $\sqrt{s}$=91.1~GeV 
to $\sqrt{s}$=175~GeV.
\item[*]The 2-dimensional moments rise much more steeply than the 
1-dimensional  moments
(Figs 3b,~3d).
\item[*]The values of $\phi_n$ obtained by fitting eq.~4 to the data in 
the region
of small $\epsilon$ ($\epsilon < 0.1$) follow the predictions 
eq.~5 qualitatively, as can be seen
in Table 3.
\item[*]In Fig.~3 it is shown that the analytically calculated factorial
moments depend sensitively on $\Lambda$. It should be noted that a similar
dependence (although weaker because of the $\Lambda$-independent fragmentation)
is observed in JETSET when varying $\Lambda$ and keeping all other
parameters constant. 
\end{itemize}
\subsection{Discussion of the QCD parameter $\gamma_0$}
The first term in the perturbative formula eq.~5 involves 
the phase space volume, the second one
depends explicitly on the parameter $\gamma_0$
(eq.~3), i.e. the QCD coupling $\alpha_s$.  
Fig.~5a summarizes the behaviour at small $\epsilon$, where
the numerical values of $\gamma_0^{eff}$ derived from the measured slopes
$\phi_n$ are given for the orders $n=2,3,4,5$.
From the present theoretical understanding,
$\gamma_0$ is expected to be independent of $n$. For example, for 
$\Lambda$=0.15~GeV and $n_f$=3 ($\Theta=25^o$, $Q\sim P\Theta$)  
eqs.~13 and 14 give the numerical value
$\alpha_s$=0.143 and hence from eq.~3 the value $\gamma_0$=0.523. 
This is indicated as horizontal line in Fig.~5, where also the lines for
$\Lambda$=0.01~GeV and $\Lambda$=0.8~GeV
are given for comparison. The average 
measured values of $\gamma_0^{eff}$ are of the same order as the expectation.
The $n$-dependence observed, however, is not described by the calculations.
The measured values of $\gamma^{eff}_0$ agree, however, extremely well with the
corresponding values obtained from JETSET, as can be seen in Fig.~5a.
\newpage
\begin{center}
{\bf Table 3: Comparison of measured and predicted slopes $\phi_n$} \\
the errors were obtained by adding 
statistical and systematic errors quadratically \\ 
\vspace{0.8mm}
%\begin{footnotesize}
\begin{tabular} {|c|c|c|c|c|}
\hline
{\bf 1-dimensional case} & {\bf n=2} & {\bf n=3} & {\bf n=4} & {\bf n=5} \\
\hline & & & & \\ 
data    & 0.38 $\pm$ 0.006& 1.04 $\pm$ 0.02& 1.87 $\pm$ 0.02& 2.78 $\pm$ 0.03\\ 
$\Lambda=0.15$~GeV, $~n_f=5$ & 0.15 & 0.49 & 0.88 & 1.28 \\
$\Lambda=0.04$~GeV, $~n_f=5$ & 0.25 & 0.66 & 1.12 & 1.59 \\
$\Lambda=0.15$~GeV, $~n_f=3$ & 0.22 & 0.61 & 1.04 & 1.49 \\
$\Lambda=0.04$~GeV, $~n_f=3$ & 0.30 & 0.76 & 1.26 & 1.77 \\
$\Lambda=0.005$~GeV, $n_f=3$ & 0.40 & 0.93 & 1.50 & 2.07 \\
 & & & &  \\
\hline
{\bf 2-dimensional case} & {\bf n=2} & {\bf n=3} & {\bf n=4} & {\bf n=5} \\
\hline & & & & \\ 
data    & 0.93 $\pm$ 0.02& 2.62 $\pm$ 0.04& 4.77 $\pm$ 0.05& 7.15 $\pm$ 0.06\\ 
$\Lambda=0.15$~GeV, $~n_f=5$ & 1.15 & 2.49 & 3.88 & 5.28 \\
$\Lambda=0.04$~GeV, $~n_f=5$ & 1.25 & 2.66 & 4.12 & 5.59 \\
$\Lambda=0.15$~GeV, $~n_f=3$ & 1.22 & 2.61 & 4.04 & 5.49 \\
$\Lambda=0.04$~GeV, $~n_f=3$ & 1.30 & 2.76 & 4.26 & 5.77 \\
$\Lambda=0.005$~GeV, $n_f=3$ & 1.40 & 2.93 & 4.50 & 6.07 \\
 & & & &  \\
\hline
\end{tabular}
%\end{footnotesize}
\end{center}

\begin{figure}
\mbox{\epsfig{file=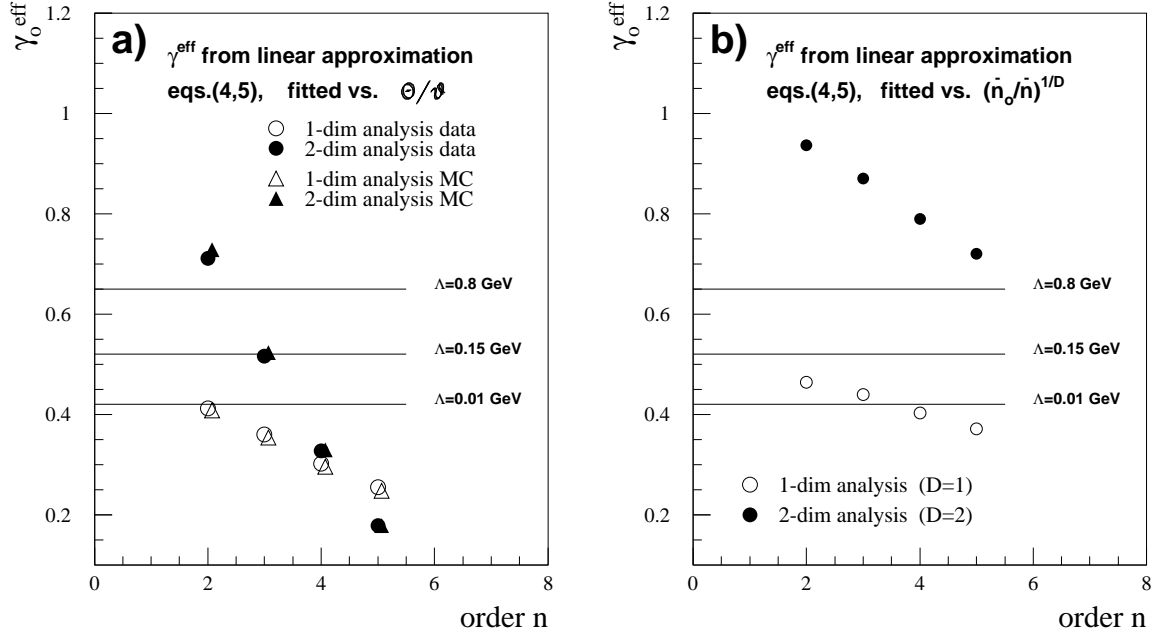,width=17cm}}
\vspace{-1.9cm}
\caption[]{
\footnotesize 
Values of $\gamma^{eff}_0$ obtained from fitting the linear approximation 
eq.~5 {\bf a)} vs. $\frac{\Theta}{\vartheta}$  and {\bf b)} vs. 
$\frac{\bar{n_0}}{\bar{n}}$
(explanation in section 4.5)
the 1-dimensional case (open circles) and the 2-dimensional case 
(full circles), for
orders $n=2,3,4,5$. In {\bf a)}, the measured values of $\gamma^{eff}_{0}$ 
are also compared 
with those obtained from JETSET at generator level: open triangles 
(1-dimension) and full triangles (2-dimensions).}
\end{figure}

\subsection{Attempts for improvement}
One of the shortcomings of the present calculations is the lack of
energy-momentum conservation.
There exist two attempts for improvement.

Firstly, in ref.~\cite{dremin}, Modified Leading Log Approximation (MLLA) 
corrections have been calculated for the 
intermittency exponents $\phi_n$. 
An order dependent correction for $\gamma_0$ has been 
proposed, 
leading to a correction to 
$\gamma_0$ amounting to only a few percent for all orders $n$ = 2 to 5. 
The deviations observed in Fig.~5 are much larger.

In a second attempt, Meunier and Peschanski~\cite{men} introduced 
energy conservation terms explicitly. This leads, however, to even 
smaller predicted slopes $\phi_n$ and consequently larger values of 
$\gamma_0$, 
increasing the discrepancies shown in Table 3 and Fig.~5.
No angular recoil effects were included in these calculations.

Recently Meunier~\cite{menpriv} proposed to use, 
instead of the evolution variable 
$\epsilon = {\ln(\Theta/\vartheta)}/{\ln(P\Theta/\Lambda)}$, the
variable $\epsilon = {\ln((\bar{n}_0/\bar{n})^{1/D})}/
{\ln(P\Theta/\Lambda)}$,
where $\bar{n}_o$ and $\bar{n}$ are the mean multiplicities in the first
$\epsilon$ bin ($\vartheta = \Theta$) and in the $\epsilon(\vartheta)$ bins
respectively. Using this new variable, the discrepancies of the
1-dimensional factorial moments observed 
so far are reduced by almost a factor 2 -- see Fig.~5b -- and the $n$-dependence 
is less strong. 
The discrepancy between the 1- and 2-dimensional moments, however, is increased
(Fig.~5b).
Whether the use of the evolution variable $\frac{\bar{n}_o}{\bar{n}}$
is more suitable than the angular evolution variable 
$\Theta/\vartheta$, which is indicated only in the 1-dimensional 
case, must
still remain open. 

Another question concerns the range of validity of the LPHD hypothesis, which
can be studied only by using Monte Carlo simulations at both partonic
and hadronic levels. Different Monte Carlo models \cite{kittel} or 
different choices of the
cut-off parameter $Q_o$ at which the parton cascade is ``terminated'',
even in a moderate interval (0.3 - 0.6 GeV),
lead to different answers~\cite{ochs,wien3,chl}. 
In the strict sense LPHD demands a low cut-off scale ($Q_0 \approx$ 0.2-0.3~GeV)
\cite{lhd1,ochspriv}.
In a JETSET study of the partonic state with $\Lambda$=0.15~GeV and 
$Q_0$=0.33~GeV a steeper rise of the moments than that of the 
hadron state 
is observed at small $\epsilon$ thus even increasing the 
discrepancy with the analytical predictions. These studies and the results
of \cite{kittel} indicate that even a possible violation of LPHD 
might not be the
reason for the observed discrepancies.

Fig.~1 also shows that shape distortions due to resonance decay, 
although significant, are much smaller than 
the discrepancies between data and theoretical predictions.
Similarily a slightly steeper rise of moments is also observed in Monte 
Carlo studies when replacing the sphericity axis by the "true" $q\bar{q}$ 
axis and excluding initial heavy flavour production. These effects, 
however, are smaller than that caused by inhibiting resonance decay 
(see Fig.~1c,d).

This discussion suggest that the analytical calculations need to be 
improved beyond the above attempts. \underline{Only after improving the 
perturbative calculations} does one have a better handle to estimate 
how far nonperturbative effects are spoiling the agreement with the
data. The importance of including angular recoil effects into the parton 
cascade, as it is also stressed in \cite{ochs}, is intuitively evident when
analysing angular dependent functions.

\section{Summary and outlook} 

Experimental data on
multiplicity fluctuations in one- and two- dimensional angular intervals
in $e^+e^-$ annihilations into hadrons at $\sqrt{s} = 91.1$~GeV
and $\sqrt{s} \sim 175$~GeV
collected by the DELPHI detector 
have been compared with first order analytical calculations of the DLA and 
MLLA of perturbative QCD. Some general features of the calculations are confirmed by the data: the factorial moments rise approximately linearly for large
angles (as expected from the multifractal nature of the parton shower) and
level off at smaller angles; the dimensional-, order- and energy
dependences are
met qualitatively.

At the quantitative level, however, large deviations are observed: 
the cumulants are far off the predictions; the factorial moments level off with 
substantially smaller radii; even by reducing the QCD parameters $\Lambda$ 
and/or $n_f$, the analytical calculations are not able to describe
simultaneously the factorial moments at \underline{all} orders $n=2,3,4,5$ 
and at different dimensionalities (1- and 2-dimensions). 
Thus an evaluation of
QCD parameters from the data is not possible at present.
From Monte Carlo studies there are indications that possible violations
of LPHD are not responsible for these discrepancies.

Therefore these shortcomings  
are probably mainly due to the high energy approximation inherent in
the DLA
(which is most responsible for the extreme failure of 
calculations using cumulants). 
Available MLLA calculations cannot substantially improve on the DLA.
To match the data at presently available energies, improvements such as 
the inclusion of full energy-momentum conservation are needed.

Similar conclusions have been obtained by a parallel 
one-dimensional study~\cite{kittel}.
More checks on refined predictions are desirable in the future.

\begin{flushleft} {\large\bf Acknowledgements} \end{flushleft} \vspace{-0.3cm} 
We thank W.
Kittel, P. Lipa, J.-L. Meunier, W.Ochs, R. Peschanski 
and J.Wosiek for valuable discussions and stimulation.

We are greatly indebted to our technical collaborators and to the funding
agencies for their support in building and operating the DELPHI detector, and 
to the members of the CERN-SL Division for the excellent performance of the
LEP collider.
 
%\begin{flushleft} {\large{\bf References}} \end{flushleft}
%\vspace{-1.9cm}

\end{document}